\documentclass{article}
\begin{document}
 
\title{ Exact Description of Decoherence in Optical Cavities}
\author{ K. M. Fonseca 
Romero$^{(1)}$\thanks{  E--mail: karenf@ciencias.unal.edu.co }
{ } and M. C. Nemes$^{(2,3)}$\thanks{  E--mail: carolina@fisica.ufmg.br }}

\maketitle

\begin{center}
{$^{(1)}$ Departamento de F\'{\i}sica, Facultad de Ciencias, \\
\small \it Universidad Nacional, Ciudad Universitaria,
Bogot\'a, Colombia}
  
{$^{(2)}$ Departamento de F\'{\i}sica--Matem\'atica, Instituto de
 F\'\i sica, \\
\small \it Universidade de S\~ao Paulo, C.P. 66318, 05315-970
 S\~ao Paulo, S.P., Brazil}

{$^{(3)}$ Departamento de F\'\i sica, ICEX,\\
\small \it Universidade Federal de
 Minas Gerais, C.P. 702, 30161-970 Belo Horizonte,
 M.G., Brazil}

\end{center}

\begin{abstract}
The exact reduced dynamics for the independent oscillator model in the 
RWA approximation at zero and finite temperatures is derived. It is 
shown that the information about the interaction and the environment is 
encapsulated into three time dependent coefficients of the master 
equation, one of which vanishes in the zero temperature case. In 
currently used optical cavities all the information about the field 
dynamics is contained into {\it two} (or three) experimentally accesible 
and physically meaningful real functions of time. From the phenomenological
point of view it suffices  then to carefully measure two ({\it three}) 
adequate observables in order to map the evolution of any initial condition, 
as shown with several examples:  (generalized) coherent states, Fock states, 
Schr\"odinger cat states, and squeezed states.
\end{abstract}
 

\section{Introduction}
 
Measuring the time development of the entanglement process of a system 
coupled to its environment is a most remarkable achievement and a 
challenging goal. The reason for this is that the entanglement process 
is a unique and typical quantum feature. Several attemps, both on the 
theoretical as well as on the experimental side have been recently 
made\cite{theor,theor2}. In the particular case of high Q optical 
cavities a direct measure of the decoherence process has been given and
suggestions of experiments with essentially the same set up have been 
made on how to directly measure the Wigner function of the initially 
correlated field produced in the cavity\cite{Haroche,Lutterbach}.
We show that all necessary information to 
construct the time development of any Wigner function (or any system's 
density operator) can be obtained by a precise measure of three 
quantities as function of time: the average photon number and two 
orthogonal field quadratures.
 
Recently the exact master equation for quantum Brownian motion in a 
general environment has been derived using both path integral 
techniques\cite{Hu} and the tracing of the evolution equation for the 
Wigner function\cite{Habib,Halliwell}. We closely follow the later 
approach to derive the exact master equation for the oscillator 
independent model in the RWA at zero temperature and with a 
factorized initial condition. The hamiltonian of the model is 
\begin{equation}
\label{Eq:RWAHamiltonian}
H=\hbar\omega(a^\dagger a + 1/2) + 
\hbar \sum_k \omega_k(a^\dagger_k a_k + 1/2) +
\hbar \sum_k c_k (a^\dagger a_k + a^\dagger_k a).
\end{equation}
Here we present the solution of the initial value problem, which have
been solved in the Heisenberg picture in Refs. \cite{Hope,Savage}. The 
solution allows for an easy visualization in contrast to the model 
without the RWA approximation. Besides its intrinsic interest as an 
exactly soluble model, the hamiltonian (\ref{Eq:RWAHamiltonian}) may be useful
in treating leaking Bose-Einstein condensates \cite{Moy}, in materials with
modified dispersion relations\cite{Bay}, or in any case of non-ohmic strength
function, where the Born-Markov approximation is not adequate\cite{Moy}.

We assume that at $t=0$ the total density operator is given by
\begin{equation} \label{Eq:InitialRho}
\rho(0) = \rho_S(0) \otimes 
\prod_k \frac{e^{-\beta \hbar \omega_k a_k^\dagger a}}
{{\rm Tr }e^{-\beta \hbar \omega_k a_k^\dagger a}}
\stackrel{\beta \rightarrow \infty}{\longrightarrow}
\rho_S(0) \otimes \prod_k |0_k\rangle \langle0_k|, 
\end{equation}
where the subscript $S$ for system refers to the main oscillator. The 
bath, i.e. the set of oscillators labelled by $k$, is initially in 
thermal equilibrium at inverse temperature $\beta$. At zero temperature, the 
tensor product of the vacuum of the 
main oscillator with the vacuum of the set of oscillators is the ground 
state of (\ref{Eq:RWAHamiltonian}). 

\section{The Exact Master Equation}
It is well known that for 
quadratic hamiltonians the Wigner function satisfies the classical 
Liouville equation. To obtain the classical hamiltonian corresponding 
to (\ref{Eq:RWAHamiltonian}), one uses the correspondence rule
$a_\mu^{(\dagger)} \rightarrow  \alpha_\mu^{(*)}$, where 
\begin{equation}
\alpha_\mu^{(*)} = \sqrt{\frac{m_\mu \omega_\mu}{2 \hbar}} q_\mu
+ (-) \frac{i}{\sqrt{2\hbar m_\mu \omega_\mu}}p_\mu,
\end{equation}
and discards the zero energy contributions. We use $\mu=0,1,2,..$, 
$k=1,2,..$ and $a_0\equiv a$, $\omega_0 = \omega$. In general greek
subindices denote non negative integers while latin sub indices denote
positive integers. Using these 
conventions, the classical Liouville equation 
\begin{equation}
\label{Eq:Liouville}
\frac{\partial W(\alpha_\mu,\alpha_\mu^*,t)}{\partial t} =
\frac{1}{i\hbar} \sum_{\mu} \frac{\partial H}{\partial \alpha_\mu}
\frac{\partial W}{\partial \alpha_\mu^*} - 
\frac{1}{i\hbar} \sum_{\mu} \frac{\partial H}{\partial \alpha_\mu^*}
\frac{\partial W}{\partial \alpha_\mu},
\end{equation}
where $W(\alpha_\mu,\alpha_\mu^*,t)$ is the Wigner function in the
quantum case and the probability density function in the classical case, 
can be written as
\begin{eqnarray} \nonumber
\frac{\partial W(\alpha_\mu,\alpha_\mu^*,t)}{\partial t} & = &
-i\omega \alpha^* \frac{\partial W}{\partial \alpha^*}
+i\omega \alpha \frac{\partial W}{\partial \alpha}
-i\sum_k \omega_k \alpha_k^* \frac{\partial W}{\partial \alpha_k^*}
+i\sum_k \omega_k \alpha_k \frac{\partial W}{\partial \alpha_k}\\
\label{Eq:WignerTotal} & &
-i\sum_k c_k \alpha_k^* \frac{\partial W}{\partial \alpha^*}
+i\sum_k c_k \alpha_k \frac{\partial W}{\partial \alpha}
-i\sum_k c_k \alpha^* \frac{\partial W}{\partial \alpha_k^*}
+i\sum_k c_k \alpha \frac{\partial W}{\partial \alpha_k}.
\end{eqnarray}
The initial condition Eq. (\ref{Eq:InitialRho}) in the language of
Wigner functions is 
\begin{equation}
\label{W0}
W(\alpha_\mu,\alpha_\mu^*,t=0) = 
W^0(\alpha_\mu,\alpha_\mu^*) =
W^0_S(\alpha,\alpha^*) W^0_B(\alpha_k,\alpha_k^*)=
W^0_S(\alpha,\alpha^*) \prod_k N_k 
{\rm e}^{-2\tanh (\hbar\omega_k\beta)\alpha_k \alpha_k^*},
\end{equation}
where the $N_k$ are normalization constants.
Integrating Eq.(\ref{Eq:WignerTotal}) over the bath variables we get 
\begin{equation}
\frac{\partial \widetilde{W}(\alpha,\alpha^*,t)}{\partial t} =
-i\omega \alpha^* \frac{\partial \widetilde{W}}{\partial \alpha^*}
+i\omega \alpha \frac{\partial \widetilde{W}}{\partial \alpha}
-i \frac{\partial G^*}{\partial \alpha^*}
+i \frac{\partial G}{\partial \alpha},
\end{equation}
with
\begin{equation}
\widetilde{W}(\alpha,\alpha^*,t) = \int \left(
\prod_k d\alpha_k d\alpha_k^* \right) 
W(\alpha_\mu,\alpha_\mu^*,t),
\end{equation} 
and  
\begin{equation}
G(\alpha,\alpha^*,t) = \int \left( 
\prod_k d\alpha_k d\alpha_k^* \right)
\sum_k c_k \alpha_k W(\alpha_\mu,\alpha_\mu^*).
\end{equation}
As in \cite{Halliwell} it is easy to show that $G(\alpha,\alpha^*,t)$ 
can be written in terms of $\widetilde{W}$. We notice that for quadratic 
hamiltonians $W(\alpha_\mu,\alpha_\mu^*,t) = W^0 (\alpha_\mu(-t), 
\alpha_\mu^*(-t))$,  where $\alpha_\mu(t)$ is the solution of the 
classical equations of 
motion. If we define $\vec{\alpha}(t)=(\alpha_0(t), \alpha_1(t), \ldots)$ 
and denote its transpose by $\vec{\alpha}^T(t)$, we have
\begin{equation} \label{Eq:Solution}
\vec{\alpha}^T(t) = 
U^\dagger \Delta(t) U \vec{\alpha}^T(0),
\hspace{.5cm} \vec{\alpha}^* (t) = (\vec{\alpha}(t))^*,
\end{equation}
where $U$ and $\Delta$ are unitary, and $\Delta$ is diagonal.
Taking the Fourier transform of $G$, and changing variables from 
$\{\alpha_\mu(-t),\alpha_\mu^*(-t)\}$ to $\{\alpha_\mu(0), 
\alpha_\mu^*(0)\}$, with unit Jacobian, we obtain
\begin{eqnarray} \nonumber
G(\kappa,\kappa') =& &  \int \prod_\mu d\alpha_\mu(0) d\alpha_\mu^*(0) 
{\rm e}^{i\kappa\sum_\nu p_\nu(t) \alpha_\nu (0)} 
{\rm e}^{i\kappa'\sum_\nu p_\nu^*(t) \alpha_\nu^* (0)} 
\\ \label{Eq:Gdek}&&
\times \sum_\nu q_\nu (t) \alpha_\nu (0)
W_S^0(\alpha^{(*)}(0)) W_B^0(\alpha_k^{(*)}(0)),
\end{eqnarray}
with $\{p_\nu,p_\nu^*,q_\nu \}$ time dependent parameters.
From Eq.(\ref{Eq:Gdek}) it is easy to see that the multiplication by 
$\alpha_0 = \alpha$ is equivalent to a derivation with respect to $k$, 
plus terms corresponding to multiplication by $\alpha_k$, up to time 
dependent coefficients. These last terms, as can be seen from 
(\ref{W0}), correspond to derivations w.r.t. $\alpha_k^*$, 
which in turn, are equivalent to multiplication by $k'$, as shows a 
simple integration by parts. Taking the inverse Fourier transform we 
obtain a multiplication by $\alpha$ and a derivation w.r.t. 
$\alpha^*$. Thus, observing that the Fourier transform of 
$\widetilde{W} (\alpha,\alpha^*)$, $\widetilde{W} (\kappa,\kappa')$ 
is given by
\begin{eqnarray}\nonumber
\widetilde{W} (\kappa,\kappa') = &&
\int \prod_\mu d\alpha_\mu(0) d\alpha_\mu^*(0) 
{\rm e}^{i\kappa\sum_\nu p_\nu(t) \alpha_\nu (0)} 
{\rm e}^{i\kappa'\sum_\nu p_\nu^*(t) \alpha_\nu^* (0)} 
\\ \nonumber&&
\times W_S^0(\alpha^{(*)}(0)) W_B^0(\alpha_k^{(*)}(0)),
\end{eqnarray}
we obtain
\begin{eqnarray} \nonumber
i\frac{\partial G}{\partial \alpha} - 
i\frac{\partial G^*}{\partial \alpha^*}
& = & iY \frac{\partial}{\partial \alpha}
\left( \alpha \widetilde{W} \right)
-iY^* \frac{\partial}{\partial \alpha^*}
\left( \alpha^* \widetilde{W} \right)\\ & &
+(iZ-iZ^*) \frac{\partial^2 \widetilde{W}}
{\partial\alpha\partial\alpha^*},
\end{eqnarray}
with time dependent functions $Y, Z$. Therefore the Wigner equation 
can be written as
\begin{eqnarray}\nonumber
\frac{\partial \widetilde{W}(\alpha,\alpha^*,t)}{\partial t} & = &
-i(\omega+\delta) 
\left( \alpha^* \frac{\partial \widetilde{W}}{\partial \alpha^*}
- \alpha \frac{\partial \widetilde{W}}{\partial \alpha}\right)
+2\lambda \widetilde{W}\\ \label{Eq:WignerEq} & & 
+\lambda \left(\alpha^*\frac{\partial \widetilde{W}}{\partial \alpha^*}
+ \alpha\frac{\partial \widetilde{W}}{\partial \alpha}\right)
+\lambda' \frac{\partial^2 \widetilde{W}}
{\partial\alpha\partial\alpha^*},
\end{eqnarray}
where we have set $iY = \lambda + i\delta$ and $iZ-iZ^* = \lambda'$. 
All of $\lambda, \delta$ and $\lambda'$ are real functions.  By 
comparing the system of equations  found from both (\ref{Eq:WignerTotal}) 
and (\ref{Eq:WignerEq}) we get 
\begin{eqnarray} 
(\lambda+i\delta)\langle \alpha \rangle & = & i
\sum_k c_k \langle \alpha_k \rangle, \\ 
(\lambda+i\delta)\langle \alpha^2 \rangle & = & i
\sum_k c_k \langle \alpha \alpha_k \rangle, \\
-2\lambda' \langle \alpha \alpha^*\rangle & = & 
i \sum_k c_k \langle \alpha \alpha_k^*\rangle 
-i \sum_k c_k \langle \alpha^* \alpha_k \rangle.
\end{eqnarray}
We know that the solution of the Heisenberg equations can be written 
as follows
\begin{eqnarray} \label{aoft}
a(t) & = & \eta(t) a(0) + \sum_k \gamma_k(t) a_k(0), \\ \label{akoft}
a_k(t) & = & \eta_k(t) a(t) + \sum_l \gamma_{kl} a_l(0).
\end{eqnarray}
Using the above solution of the Heisenberg equations, and the fact that  
all of the first and second (symmetric) moments involving bath operators
are zero, with the exception of $\langle \{a_k^\dagger, a_k\} \rangle =
2n_k+1=2\coth (\hbar\omega_k\beta/2)$, we obtain
\begin{eqnarray}\label{Eq:Sol1}
\lambda(t)+i\delta(t) & = & i \sum_k c_k \eta_k(t)\\
\label{Eq:Sol2}
\lambda'(t) & = & \sum_{kl} c_k (n_k(\beta)+ \frac{1}{2})
(i\gamma_l\gamma_{kl}^*-i\gamma_l^*\gamma_{kl}).
\end{eqnarray}
For reasons that will be clear soon, we write the diffusion coefficient
$\lambda'(t)$ as $\lambda(t)+\epsilon(t,\beta)$, with $\lim_{\beta
\rightarrow \infty}\epsilon(t,\beta)=0$, as shown in section \ref{DMEC}. 
At zero
temperature, since the tensor product of vacua is the ground state of 
(\ref{Eq:RWAHamiltonian}), the corresponding (reduced) Wigner function 
$W_S(\alpha,\alpha^*)$ should be a stationary solution of the Wigner 
equation. When this condition is applied to Eq.(\ref{Eq:WignerEq}), 
we obtain $\lambda = \lambda'$. If they were not equal it would imply 
the non existence of an exact master equation, as seems to be claimed
in Ref. \cite{Moy}. However, this is not the 
case, as we show next.
 
It is not hard to show that the operator equation for the system's 
reduced density operator equivalent to the Wigner equation 
(\ref{Eq:WignerEq}) is
\begin{equation} \label{Eq:Master}
\frac{d\rho}{dt} = 
\frac{1}{i\hbar}\left[\hbar(\omega+\delta)a^\dagger a,\rho \right]
+(\lambda+\epsilon)(2 a\bullet a^\dagger-a^\dagger a\bullet 
-\bullet a^\dagger a)\rho +\epsilon(2 a^\dagger\bullet a-a 
a^\dagger\bullet 
-\bullet a a^\dagger)\rho=
{\cal L}(t)\rho(t),
\end{equation}
where the usual dot superoperator convention has been 
used. The usual Born-Markov RWA master equation is of 
this form with constant coefficients\cite{Louisell}. Some results can 
be obtained at once from (\ref{Eq:Master}): premultiplying by $a$ and 
taking the trace we get
\begin{equation}
\frac{d}{dt}\langle a \rangle = \frac{d\alpha}{dt} =
(-i(\omega+\delta)-\lambda)\alpha,
\end{equation}
which can be immediately solved to give
\begin{equation} \label{Eq:Alphat}
\alpha(t)=\exp(-i\Omega(t)-\Lambda(t))\alpha(0),
\end{equation}
with
\begin{equation} \label{Eq:Defs}
\Omega(t)=\int_0^t d\tau(\omega+\delta)(\tau),
\hspace{0.5cm}
\Lambda(t) =\int_0^t d\tau \lambda(\tau).
\end{equation}
Note that this result is {\it independent} of $\epsilon$, i.e., it does 
not depend on the temperature. 
Premultiplying (\ref{Eq:Master}) by $a^\dagger a$, and taking the trace 
we get the following differential equation
\begin{equation} \label{eqforepsilon}
\frac{d}{dt}\langle a^\dagger a \rangle (t) =
-2 \lambda \langle a^\dagger a \rangle (t) + 2 \epsilon
\end{equation}
with the solution
\begin{equation} \label{meanadaggera}
\langle a^\dagger a\rangle (t)= \exp(-2\Lambda(t))
\langle a^\dagger a\rangle (0)+{\cal N}(t)=
\langle a^\dagger a\rangle (t;\beta\rightarrow\infty)+
2\exp(-2\Lambda(t))\int_0^t d\tau \epsilon(\tau)\exp(2\Lambda(\tau)),
\end{equation}
where it is evident that ${\cal N}(t)$ vanishes in the zero temperature 
limit. Contrary to the exact equations found in \cite{Hu,Habib,Halliwell}
We thus have the following interpretations for the real functions 
that appear in the master equation: $\delta(t)$ is the instantaneous frequency
shift, $\lambda(t)$ is the instantaneous energy rate of change at zero 
temperature and $\epsilon(t)$ is the instantaneous energy rate of change at 
finite temperature but with the system in the vacuum state. Moreover
${\cal N}(t)$, which is related to both $\epsilon(t)$ and $\delta(t)$ is the 
mean number of excitations when the initial state was the ground state.   
  
\section{The evolution superoperator and some initial states}
 
We can use Lie algebraic methods\cite{Borges} to find the evolution 
superoperator ${\cal U}$. Indeed, we can verify that the 
superoperators ${\cal M} = a^\dagger a\bullet$, ${\cal P} = 
\bullet a^\dagger a$, ${\cal J} = a\bullet a^\dagger$ and 
${\cal R} = a^\dagger\bullet a$ form an 
algebra,
\begin{equation}\label{Eq:Commutation}
[{\cal M},{\cal P}]=0, \quad [{\cal M},{\cal J}]
= - {\cal J} = [{\cal P},{\cal J}],\quad
[{\cal M},{\cal R}]
= - {\cal R} = [{\cal P},{\cal R}].
\end{equation}
Thus, we can assume that 
${\cal U}(t) = v {\rm e}^{w{\cal R}} {\rm e}^{x{\cal M}} {\rm e}^{y{\cal 
P}} 
{\rm e}^{z{\cal J}}$.
Deriving this expression, using the formula $ \exp(xA) B \exp(-xA) 
= B+x[A,B]+x^2 [B,[B,A]]/2! +...$ and the commutation relations 
(\ref{Eq:Commutation}), comparing coefficients in the equation 
$d{\cal U}/dt$ = ${\cal L}(t) {\cal U}(t)$, and solving the resulting 
differential equations we obtain
\begin{eqnarray} \nonumber
v(t) & = & \frac{1}{1+{\cal N}(t)}, \quad
w(t)  =  \frac{{\cal N}(t)}{1+{\cal N}(t)}, \quad
x(t)  =  -i\Omega(t)-\Lambda(t)-\frac{1}{2}\ln (1+{\cal N}(t)) =  
y^*(t), \\ \label{Ucoefs}
& & z(t)  =  1-\frac{\exp(-2 \Lambda(t))}{1+{\cal N}(t)}.
\end{eqnarray}
 
Let us suppose that $\rho$ satisfies the equation $d\rho/dt = 
{\cal L} (X_i\bullet,\bullet X_i;t) \rho(t)$, where the $X_i$ are 
operators (notice that ${\cal L}$ is a general linear superoperator), 
and that $\rho$ can be written as $U\rho'U^{-1}$. Then, $\rho'$ satisfy 
the equation $d\rho'/dt = {\cal L}' \rho'(t)$. If, moreover, we
choose $U=\exp(\sigma a^\dagger -\sigma^* a)=D(\sigma)$, the 
displacement operator, and ${\cal L}$ is that of the RWA, then
\begin{equation}
{\cal L}'(t)  =  {\cal L}(t) + 
\left( (i\dot{\Omega}+\lambda)\sigma+\frac{d\sigma}{dt}\right)
(a^\dagger\bullet-\bullet a^\dagger)  -
\left( (-i\dot{\Omega}+\lambda)\sigma^*+\frac{d\sigma^*}{dt}\right)
(a\bullet-\bullet a).
\end{equation}
It is easy to see that if
\begin{equation} \label{sigmaoft}
\sigma(t)=\sigma(0)\exp(-i\Omega(t)-\Lambda(t)), 
\hspace{0.5cm} \sigma^*(t) = (\sigma(t))^*,
\end{equation}
then both $\rho$ and $\rho'$ satisfy the same equation. That is, we 
have shown that $D(\sigma(t))\rho(t)D^\dagger(\sigma(t))$ satisfies Eq.
(\ref{Eq:Master}) whenever $\rho(t)$ does the same. We remark that 
this result does {\it not} depend on the temperature of the bath.

We now turn to the evaluation of the density matrix evolved with the 
superoperator found above. We chose initial states relevant from the
point of view of quantum optics. As a first initial state we choose the
system's ground state, $\rho(0)=\mid 0\rangle \langle 0\mid$. Since
\begin{eqnarray}
{\rm e}^{x{\cal M}}\mid 0\rangle \langle 0\mid &= &{\rm e}^{y{\cal P}} 
\mid 0\rangle \langle 0\mid = {\rm e}^{z{\cal J}}\mid 0\rangle \langle 
0\mid = \mid 0\rangle \langle 0\mid,
\quad {\rm and} \\
{\rm e}^{x{\cal R}}\mid 0\rangle \langle 0\mid & = & 
\sum_0^\infty \frac{x^n}{n!} (a^\dagger)^n
\mid 0\rangle \langle 0\mid a^n= \sum_0^\infty x^n 
\mid n\rangle \langle n\mid,
\end{eqnarray}
we have
\begin{equation}
\rho(t) = {\cal U}(t) \mid 0\rangle \langle 0\mid =
\sum_0^\infty \frac{1}{1+{\cal N}}\left(1+1/{\cal N}\right)^n
\mid n\rangle \langle n\mid = \sum_0^\infty P_n(t)
\mid n\rangle \langle n\mid.
\end{equation}
The above formula displays the so called decomposition in natural 
orbits\cite{Nemes} where the quantities $P_n$ can be directly 
interpreted as probabilities. We can write the evolved density matrix
in the alternative form $\rho(t) = \exp((1+1/{\cal N})a^\dagger a)/ 
(1+{\cal N})$, which is the form of an instantaneous thermal density
matrix, with ${\cal N}(t)=<a^\dagger a>(t)$. Had we chosen an initial
thermal state, with mean number of excitations $\bar{n}(0)$, the density
matrix would have remained a thermal state, but now $M(t) = 
\bar{n}(0)\exp(-2\Lambda(t))+{\cal N}(t)$. If we use the instantaneous
oscillator frequency $\omega'=\omega+\delta$,
it is possible to define an instantaneous temperature through the 
relation $T(t)= \hbar(\omega+\delta)/(k_B \ln(1+1/{\cal N}))$, with 
$k_B$ the Boltzmann constant. Moreover, we have obtained  a physical
interpretation for the quantity ${\cal N}(t)$: it is the mean number of
excitations of the main oscillator at time $t$ when it was initially
prepared in its ground state.

To calculate the density matrix for an initial Fock state it is better to
write the evolution superoperator in the form ${\cal U}(t) = v 
\exp (w{\cal R})  \exp(z'{\cal J}) \exp(x{\cal M}) \exp(y{\cal P})$, 
where $w,x,y$ are given in (\ref{Ucoefs}), and $z'(t) = (1+{\cal 
N})\exp(2\Lambda)$. We use
\begin{eqnarray} \nonumber
e^{z{\cal J}} e^{x{\cal M}}e^{y{\cal P}} \mid m\rangle \langle m\mid & = 
& \sum_{k=0}^{m} (e^{x+y})^{m-k}(ze^{x+y})^{k} \frac{m!}{(m-k)!k!} 
\mid m-k\rangle \langle m-k\mid \\ & = &
\sum_{k=0}^{m}  \frac{m!}{(m-k)!k!} (e^{x+y})^{k} (ze^{x+y})^{m-k}
\mid k\rangle \langle k\mid,
\quad and\\
e^{u{\cal R}}  \mid m\rangle \langle m\mid & = &
\sum_{k=0}^{\infty} \frac{m! u^k}{(m-k)!k!} 
\mid k\rangle \langle k\mid,
\end{eqnarray}
to see that the density matrix at time $t$ is given by $\rho(t) = 
\sum_{k=0}^\infty P_{m,s}(t) \mid s\rangle \langle s\mid$, with
\begin{equation} \label{probsFock}
P_{m,s}(t) = \frac{e^{-2m\Lambda}}{(1+{\cal N})^{m+1}}
\frac{m!}{s!} \sum_{k=0}^{{\rm min}(m,s)} 
\frac{([1+{\cal N}]e^{2\Lambda}-1)^k}{(m-k)!(s-k)!}
\left( 1+\frac{1}{{\cal N}}\right)^{s-k}.
\end{equation}
Since the former density matrix have been expressed in terms of natural 
orbits, the quantities $P_{m,s}(t) $ are readily interpreted as
probabilities. The transformation property discussed above allows us to 
write the evolution of an initial generalized coherent state $\mid \sigma
m \rangle = D(\sigma) \mid m\rangle $, where $D$ is the displacement
operator and $\mid n\rangle$ the n-th number state. We have
${\cal U}(t)\mid \sigma_0 m\rangle \langle \sigma_0 m\mid =  
\sum_{k=0}^\infty P_{m,s}(t) \mid \sigma(t) s\rangle \langle \sigma(t) 
s\mid$, with $ P_{m,s}(t)$ given by (\ref{probsFock}) and $\sigma(t)$ by 
(\ref{sigmaoft}).

One interesting point to be investigated is if there exists an 
asymptotic density operator. Provided that our environment is such that 
$\lim_{t\uparrow\infty}\Lambda(t)\rightarrow\infty$ and
$\lim_{t\uparrow\infty} {\cal N}(t)= n_\infty$, the asymptotic evolution
superoperator can be written as
\begin{equation}
\lim_{t\uparrow\infty} {\cal U}(t) =
\frac{1}{1+n_\infty}
\exp (\frac{n_\infty}{1+n_\infty}{\cal R}) 
(\mid 0 \rangle \langle 0 \mid \bullet) 
\exp({\cal J}) (\bullet \mid 0 \rangle \langle 0 \mid ),
\end{equation}
which applied to a generic normalized initial density $\rho(0)$ gives
\begin{eqnarray}\nonumber
\rho_\infty & = & 
\frac{1}{1+n_\infty}
\exp({\cal J}) \rho(0)\langle 0 \mid
\exp (\frac{n_\infty}{1+n_\infty}{\cal R}) \mid 0 \rangle \langle 0 \mid 
\\ & = & 
\frac{1}{1+n_\infty}\left({\rm Tr} \rho(0)\right)
\exp [(1+\frac{1}{n_\infty})a^\dagger a]\mid 0 \rangle \langle 0 \mid 
= \frac{1}{1+n_\infty}
\exp [(1+\frac{1}{n_\infty})a^\dagger a]\mid 0 \rangle \langle 0 \mid.
\end{eqnarray}
Thus, whenever the established conditions are met, the density 
operator approaches asymptotically  to a thermal state with a mean 
number of excitations equal to that of the environment.  The existence
of a unique asymptotic density can not be taken for granted: in the 
model of decoherence without damping studied in references \cite{qdec} 
even when the coefficient of decoherence grows indefinitely with time,
the asymptotic state depends on the initial state.

The normal order characteristic functional$C^{(n)} (\xi,\xi^\dagger,t)$
given by \cite{Louisell}
\begin{equation}
C^{(n)} (\xi,\xi^\dagger,t) = 
{\rm Tr } e^{i\xi a^\dagger} e^{i\xi^* a} \rho(t) =
{\rm Tr } e^{i\xi a^\dagger} e^{i\xi^* a} {\cal U}(t) \rho(0) =
{\rm Tr } {\cal U}^\dagger(t) e^{i\xi a^\dagger} e^{i\xi^* a} \rho(0),
\end{equation}
where ${\cal U}(t)$ is the evolution superoperator and ${\cal 
U}^\dagger(t)$ its adjoint, is the generating functional of the normally
ordered moments. After a somewhat lengthy but straightforward calculation
we obtain
\begin{equation}
C^{(n)} (\xi,\xi^\dagger,t) = 
e^{-2 {\cal N}\xi xi^\dagger} 
{\rm Tr } e^{i\xi \exp(-\Lambda+i\Omega)a^\dagger}
 e^{i\xi^* \exp(-\Lambda-i\Omega) a} \rho(0).
\end{equation}
If we calculate $\langle a \rangle (t)$ and 
$\langle a\dagger a \rangle (t)$ using  
\begin{equation}
\langle a \rangle (t)= 
\frac{\partial}{\partial \xi^*}C^{(n)} 
(\xi,\xi^\dagger,t)\mid_{\xi=0=\xi^\dagger},
\quad
\langle a\dagger a \rangle (t)=
\frac{\partial^2}{\partial \xi\partial \xi^*}
C^{(n)} (\xi,\xi^\dagger,t)\mid_{\xi=0=\xi^\dagger},
\end{equation}
we arrive at the same results as before. These can be compared to those 
obtained in reference \cite{Hope}.

\section{ The Zero Temperature Limit}

The zero temperature limit has its own special interest, both as an 
approximation at low temperatures, and as the relevant case for leaking 
Bose-Einstein condensates. In this case the evolution superoperator can be
expressed as ${\cal U}(t) =  {\rm e}^{\tilde{x}{\cal M}} 
{\rm e}^{\tilde{y}{\cal P}} {\rm e}^{\tilde{z}{\cal J}}$ with 
\begin{equation} \label{Utildecoefs}
\tilde{x}(t)  =  -i\Omega(t)-\Lambda(t)=  \tilde{y}^*(t), \quad 
\tilde{z}(t)  =  1-{\exp(-2 \Lambda(t))}.
\end{equation}

Since the vacuum is solution of the master equation, the transformation
property discussed above indicates that 
\begin{equation}\label{Eq:CState}
D(\sigma(t))| 0 \rangle\langle 0| D^\dagger (\sigma(t)) = 
|\sigma(t)\rangle \langle \sigma(t)| = {\cal U}(t)
|\sigma_0\rangle \langle \sigma_0|
\end{equation}
also solves the master equation. That means that initial coherent 
states evolve preserving their coherence, not matter what the details 
of the environment. A measurement of the norm 
and the phase of an initial coherent state is enough to determine the 
functions $\Omega$ and $\Lambda$, and hence, in principle, to determine
the evolution of any other initial state.
 
The evolution of an initial Fock state is also easily calculated. 
\begin{equation} \label{Eq:FState}
{\cal U}(t) | m \rangle\langle m|
=  \sum_{k=0}^{m} p_{k,m}(t) | k\rangle\langle k| =
p_{k,m}(t) = \frac{m!}{k! (m-k)!} ({\rm e}^{-2\Lambda(t)})^k
(1-{\rm e}^{-2\Lambda(t)})^{m-k}| k\rangle\langle k| .
\end{equation}
From this solution we can generate another solution: if we have an 
initial generalized coherent state, $| m;\sigma_0 \rangle = 
D(\sigma_0) | m \rangle$, it evolves into a mixture of generalized 
coherent states, as given by 
\begin{equation} \label{Eq:GCState}
{\cal U}(t) | m;\sigma_0 \rangle\langle m;\sigma_0| = 
\sum_{k=0}^{m} p_{k,m}(t) | k;\sigma(t)\rangle\langle k;\sigma(t)|. 
\end{equation}
 
Applying the evolution operator ${\cal U}(t)$ to an initial 
density matrix element $| \sigma_0\rangle\langle {\sigma'}_0|$, one 
obtains
\begin{equation} \label{Eq:ABEvolution}
{\cal U}(t) | \sigma_0\rangle \langle {\sigma'}_0| =
\frac{\langle {\sigma'}_0|\sigma_0\rangle}
{\langle {\sigma'}(t)|\sigma(t)\rangle} 
| \sigma(t)\rangle \langle \sigma'(t)|.
\end{equation}
Notice that an heuristic argumentation also leads to 
(\ref{Eq:ABEvolution}). Indeed, since a coherent state remains 
coherent, we expect 
${\cal U}(t) | \sigma_0\rangle \langle {\sigma'}_0| =
N(t) | \sigma(t)\rangle \langle \sigma'(t)|$.
As long as the exact dynamical equation for $\rho$ preserves 
the trace, 
$ \frac{d}{dt}{\rm Tr} \rho_S(t) = {\rm Tr} (\frac{d\rho_S(t)}{dt}) =
{\rm Tr} ({\cal L}(t)\rho_S(t)) = 0$, the normalization factor $N(t)$
cannot be other than that of Eq. (\ref{Eq:ABEvolution}).  
The evolution of an initial even ($\rho_{\sigma_0e}$) or odd cat 
($\rho_{\sigma_0o}$) state can be calculated 
from (\ref{Eq:ABEvolution}) and (\ref{Eq:CState}). Indeed,
\begin{equation}
\rho_{\sigma_0e(o)} = 
N_{e(o)}(\sigma_0) 
(|\sigma_0\rangle, |-\sigma_0\rangle)
\left( \begin{array}{lr} 
1 & (-) 1 \\ (-) 1 & 1 
\end{array}\right) 
\left( \begin{array}{c} 
\langle\sigma_0| \\ \langle -\sigma_0| 
\end{array} \right),
\end{equation}
where $N_{e(o)}(\sigma_0) = 
(1+(-)\langle -\sigma_0|\sigma_0\rangle)^{-1} /2$ is a normalization 
factor, evolves as follows
\begin{equation} \label{Eq:Cat}
{\cal U}\rho_{\sigma_0e(o)} = 
N_{e(o)}(\sigma_0) (|\sigma_t\rangle, |-\sigma_t\rangle)
\left( \begin{array}{c} 
1 \hspace{0.3cm}  \frac{(-)\langle -\sigma_0|\sigma_0\rangle} 
{\langle -\sigma(t)|\sigma(t)\rangle}\\ 
\frac{(-)\langle -\sigma_0|\sigma_0\rangle} 
{\langle -\sigma(t)|\sigma(t)\rangle} \hspace{0.3cm} 1 
\end{array} \right) 
\left( \begin{array}{c} |\sigma_t\rangle \\
|-\sigma_t\rangle\end{array}\right).
\end{equation}
We can rewrite Eq. (\ref{Eq:Cat}) in a more convenient way, in terms 
of natural orbitals, as
\begin{equation}
{\cal U}(t)\rho_{\sigma_0e(o)} =
p_{e(o)}^{e(o)}(t) \rho_{\sigma(t)e(o)} +
p_{e(o)}^{o(e)}(t) \rho_{\sigma(t)o(e)},
\end{equation}
with
\begin{eqnarray} \label{Eq:pdet}
p_{e(o)}^{e(o)}(t) & = & \frac{1}{2} 
\frac{1+(-)\langle -\sigma(t)|\sigma(t)\rangle}
{1+(-) \langle -\sigma_0|\sigma_0\rangle} 
\left( 1 + \frac{\langle -\sigma_0|\sigma_0\rangle} 
{\langle -\sigma(t)|\sigma(t)\rangle}\right) \\ \label{Eq:qdet}
p_{e(o)}^{o(e)}(t) & = & \frac{1}{2} 
\frac{1-(-)\langle -\sigma(t)|\sigma(t)\rangle}
{1+(-) \langle -\sigma_0|\sigma_0\rangle} 
\left( 1 - \frac{\langle -\sigma_0|\sigma_0\rangle} 
{\langle -\sigma(t)|\sigma(t)\rangle}\right).
\end{eqnarray}
Observe that an initial cat state evolves as a mixture of even 
and odd cat states. Eq. (\ref{Eq:pdet}) gives the probability of the 
cat state of the same parity as the initial state, and Eq. 
(\ref{Eq:qdet}) the probability of the cat state of the other parity.

The evolution of an initial squeezed state is also easily computed. We
notice that the density operator for the squeezed vacuum $\rho(\zeta)$,
can be written as
\begin{equation}
\rho(\zeta)= \lim_{\gamma\rightarrow\infty}
\rho(\zeta,\gamma) =  \lim_{\gamma\rightarrow\infty}
(1-e^{-\gamma}) S(\zeta) e^{-\gamma a^\dagger a} S^\dagger(\zeta),
\end{equation}
where $S(\zeta)=\exp((\zeta (a^\dagger)^2-\zeta^\dagger a^2)/4)$. Setting
$\zeta = \xi \exp{i\phi}$, the following expressions for the second
moments of $\rho(\zeta,\gamma)$ are found
\begin{equation}
\langle (a^\dagger)^2 \rangle = \frac{e^{-i\phi}}{2} \sinh (\xi),
\hspace{0.5cm}
\langle a^2 \rangle = \frac{e^{i\phi}}{2} \sinh (\xi),
\hspace{0.5cm} 
\langle \{ a,a^\dagger\} \rangle =  \cosh (\xi).
\end{equation}
The system of equations for the second moments, 
\begin{eqnarray}
\frac{d\langle a^2\rangle}{dt} = -2 \frac{d}{dt}(i\Omega+\Lambda) 
\langle a^2 \rangle, \\
\frac{d\langle (a^\dagger)^2\rangle}{dt} = 
2 \frac{d}{dt}(i\Omega-\Lambda) \langle (a^\dagger)^2 \rangle, \\
\frac{d\langle \{a,a^\dagger\} \rangle}{dt} = -2 \frac{d\Lambda}{dt}
\langle \{ a, a^\dagger\} \rangle + \frac{d\Lambda}{dt},
\end{eqnarray}
is integrated to give
\begin{eqnarray}
\langle (a^2)^{(\dagger)}\rangle(t) = 
\exp(-(-)2i\Omega-2\Lambda)\langle (a^2)^{(\dagger)}\rangle(0),
\\
\langle \{ a, a^\dagger\} \rangle (t) =
\exp(-2\Lambda) \langle \{ a, a^\dagger\} \rangle (0)
+(1-\exp(-2\Lambda))/2.
\end{eqnarray}
From these relationships we find ${\cal U}(t) \rho(\zeta) =
\rho(\zeta(t),\gamma(t))$, with
\begin{eqnarray} \label{zetaoft}
\zeta(t) = \xi(t) \exp(i\phi(t))
\hspace{0.5cm} \phi(t)=\phi_0-2\Omega\\ \label{xioft}
\xi(t) = {\rm ArcTanh} 
(\frac{\sinh(\xi_0)}{\cosh(\xi_0)+\exp(2\Lambda)-1})\\
\label{gammaoft}
\gamma(t) = {\rm ArcCoth}
(\sqrt{e^{-4\Lambda}+2\cosh(\xi_0)e^{-2\Lambda}(1-e^{-2\Lambda})
+(1-e^{-2\Lambda})^2}).
\end{eqnarray}
Now the evolution of a general initial state $\rho(\sigma,\zeta) =
D(\sigma)\rho(\zeta)D^\dagger(\sigma)$ is
\begin{eqnarray} 
{\cal U}(t) \rho(\sigma,\zeta)& = &
\rho(\sigma(t),\zeta(t),\gamma(t))= 
D(\sigma(t))\rho(\zeta(t),\gamma(t))D^\dagger(\sigma(t)), 
\quad {\rm or} \\
{\cal U}(t) \left( |\sigma_0,\zeta_0,n=0\rangle
\langle \sigma_0,\zeta_0,n=0|\right) & = &
D(\sigma(t)) S(\zeta(t)) 
(1-e^{-\gamma(t)}) e^{-\gamma(t)a^\dagger a}
S^\dagger(\zeta(t))D^{\dagger}(\sigma(t)),
\end{eqnarray}
where $\sigma(t),\zeta(t)$ and $\gamma(t)$ are those of Eqs.
(\ref{sigmaoft}) and (\ref{zetaoft}--\ref{gammaoft}), and 
\begin{equation}
|\sigma_0,\zeta_0,n\rangle = D(\sigma(t)) S(\zeta(t)) |n\rangle.
\end{equation}
The corresponding expansion in terms of natural orbits is
\begin{eqnarray}\nonumber
{\cal U}(t) \left( |\sigma_0,\zeta_0,0\rangle
\langle \sigma_0,\zeta_0, 0| \right) & = & \sum_{n=0}^\infty
p_n(t) |\sigma (t),\zeta(t),n\rangle
\langle \sigma (t),\zeta(t),n| \\
& = & \sum_{n=0}^\infty
e^{-n\gamma(t)} (1-e^{-\gamma(t)}) |\sigma (t),\zeta(t),n\rangle
\langle \sigma (t),\zeta(t),n|.
\end{eqnarray}


\section{Determination of the Master Equation Coefficients}
\label{DMEC}
Let's return to Eq.(\ref{Eq:Solution}). If we call $w_\mu$ the 
exact eigenfrequencies of (\ref{Eq:RWAHamiltonian}), we can write
(remenber that greek indices can assume the value 0, while latin
indices do not)
\begin{equation} \label{Eq:Sol3}
a_\nu (t) = \sum_{\mu\sigma} U_{\mu\nu}^* U_{\mu\sigma}
{\rm e}^{-iw_\mu t} a_{\sigma}(0) = 
\sum_{\sigma}Z_{\nu\sigma} a_{\sigma}(0).
\end{equation}
For the sake of convenience we write
\begin{equation}
\eta = Z_{00}, \hspace{0.3cm} \gamma_k = Z_{0k}, \hspace{0.3cm}
\Delta_k = Z_{k0}, \hspace{0.3cm} \Gamma_{kl} = Z_{kl}.
\end{equation}
Using (\ref{Eq:Sol3}) in Eqs. (\ref{aoft}) and (\ref{akoft}), we obtain 
the following expressions for $\eta_k$ and $\gamma_{kl}$,
\begin{equation}
\eta_k = \frac{\Delta_k}{\eta} \hspace{.5cm}
\gamma_{kl} = \Gamma_{kl} -\frac{\Delta_k \gamma_l}{\eta}.
\end{equation}
From (\ref{Eq:Sol1}) and (\ref{Eq:Sol2}) we obtain
\begin{equation} \label{lambdas}
\lambda = - \sum_k c_k {\rm Im}(\eta_k), \hspace{.5cm}
\lambda' = - \sum_k c_k (2 n_k(\beta)+1) {\rm Im}(\beta_k) = -\sum_k c_k 
(2 n_k(\beta)+1) {\rm Im}(\sum_l \gamma_l \gamma_{kl}^*).
\end{equation}
Since $U$ is unitary we have
\begin{eqnarray}\nonumber
\sum_l \gamma_l \Gamma_{kl}^* & = & \sum_{\lambda\mu\nu} 
U_{\mu 0}^* U_{\mu \lambda} U_{\nu k} U_{\nu \lambda}^* 
{\rm e}^{-i(w_\mu-w_\nu)t} - \sum_{\mu\nu} U_{\mu 0}^* U_{\mu 0} 
U_{\nu k} U_{\nu 0}^* {\rm e}^{-i(w_\mu-w_\nu)t}
\\ \label{Eq:Sum1} & = & \sum_{\mu\nu} 
U_{\mu 0}^*  U_{\nu k} {\rm e}^{-i(w_\mu-w_\nu)t} \delta_{\mu\nu}
- \eta\Delta^*_k = \sum_{\mu} U_{\mu 0}^*  
U_{\mu k} - \eta\Delta^*_k = - \eta\Delta^*_k,
\end{eqnarray}
and
\begin{equation} \label{Eq:Sum2}
\sum_l \gamma_l \gamma_l^* = \sum_{\lambda\mu\nu}
U_{\mu 0}^* U_{\mu \lambda} U_{\nu 0} U_{\nu \lambda}^* {\rm
e}^{-i(w_\mu-w_\nu)t}-\sum_{\mu\nu}
U_{\mu 0}^* U_{\mu 0} U_{\nu 0} U_{\nu 0}^* {\rm
e}^{-i(w_\mu-w_\nu)t} = 1 - \eta \eta^*.  
\end{equation}
Using (\ref{Eq:Sum1}) and (\ref{Eq:Sum2}) we get
\begin{equation} 
\beta_k  = 
\sum_l \gamma_l \gamma_{kl}^* = \sum_l \gamma_l \Gamma_{kl}^*-
\frac{\Delta_k^*}{\eta^*}\sum_l \gamma_l \gamma_l^*
-\eta\Delta_k^*
-\frac{\Delta_k^*}{\eta^*}(1-\eta\eta^*) = -\eta_k^*.
\end{equation}
Thus we finally notice that $\lambda'(t)$ can be expressed as a sum
\begin{equation}
\lambda'(t)=-\sum_k c_k (2n_k(\beta)+1){\rm Im}(\eta_k)=
\lambda(t)-2 \sum_k c_k n_k(\beta){\rm Im}(\eta_k)=
\lambda(t)+\epsilon(t;\beta).
\end{equation}
Observe that $\lim_{\beta\rightarrow\infty} \epsilon(t;\beta)=0$.
This proves that indeed we have the equality $\lambda(t) = \lambda'(t)$
at zero temeperature, as can be seen by looking at their definitions as
given in Eq. (\ref{lambdas}).

We have seen that $\delta,\lambda$ and $\epsilon$ (or $\Omega, \Lambda$
and ${\cal N}$) are all we need to characterize the effect of the bath on 
the main oscillator. The Heisenberg equations of motion, 
\begin{eqnarray} \label{Eq:System1}
\dot{a} & = & -i\omega_0 a -i\sum_k c_k a_k,\\
\label{Eq:System2}
\dot{a_k} & = & -i\omega_k a_k -i c_k a,
\end{eqnarray}
can be solved in a number of ways. For example, an implicit method gives
\begin{equation}
a(t) = a(0) e^{-i\Omega(t)-\Lambda(t)} -i\sum_k c_k a_k(0)
\int_0^t d\tau e^{-i\omega_k (t-\tau)} e^{-i\Omega(\tau)-\Lambda(\tau)}.
\end{equation}
where $\eta(t)$ satisfies the integrodifferential equation
\begin{equation} \label{Eq:Integrodif}
\dot{\eta}+i\omega \eta + \int_0^t d\tau
\sum_k c_k^2 {\rm e}^{i\omega_k(t-\tau)} \eta(\tau)=0,
\end{equation}
subject to the initial condition $eta_0=1$. Using the $\eta_k(t)$
implicitly defined above in eq (\ref{Eq:Sol1}), and taking into
account the equation satidfied by $\eta$, we obtain $\lambda+i\delta=
-i\omega+d(\ln \eta)/dt$, or $\eta(t)=\exp(-\Lambda(t)-i\Omega(t)$. We
also find
\begin{equation}
\epsilon(t) = \frac{e^{-2\Lambda(t)}}{2}\frac{d}{dt}
\left( e^{2\Lambda(t)} \sum_k c_k^2 \mid 
\int_0^t d\tau e^{-i\omega_k (t-\tau)} e^{-i\Omega(\tau)-\Lambda(\tau)}
\mid^2 n_k (\beta) \right).
\end{equation}
Had we used normal modes wewould have arrived at the expressions
\begin{eqnarray}
\eta(t) & = & \sum_{\nu}
\frac{e^{-iw_{\nu}t}}{1+\sum_k\frac{c_k^2}{(w_\nu-\omega_k)^2}}, \quad
{\rm and}\\
\epsilon(t) & = & {\rm Re}\left(\frac{2i}{\eta(t)}\sum_{k,\nu}
\frac{n(\beta)c_k^2}{w_\nu-\omega_k}
\frac{e^{-iw_{\nu}t}}{1+\sum_l\frac{c_l^2}{(w_\nu-\omega_l)^2}}
\right).
\end{eqnarray}
We moreover would have noticed that, in order not to have inverted
oscillators the following condition has to be fulfilled, $\omega> \sum_k
c_k^2/\omega_k$.
Since (\ref{Eq:Integrodif}) can be hard to solve, methods to obtain 
approximate solutions are welcome. For instance, we can expand in 
powers of $c_k$ to second order, $\alpha(t) = \exp 
(k_0 + c k_1 + c^2 k_2)$, to obtain
\begin{equation} \label{Eq:PerturbativeSol}
-i\Omega(t)-\Lambda(t) = -i\omega_0 t -\sum_k c_k^2 \int_0^t dt_1
\int_0^t dt_2 {\rm e}^{-i(\omega_k-\omega_0) t_2}.
\end{equation}
We expect (\ref{Eq:PerturbativeSol}) to be valid for small $c_k$. It 
can be shown that this is the time dependent Born--Markov approximation, 
which is valid also for ``strong'' coupling and short 
times\cite{Duarte}.

\section{Experimental Characterization}
 
We observe that the calculation of the mean energy, and of the entropy 
of the above examples of initial conditions only require the knowledge 
of $\Lambda(t)$. However, if for example one wants to measure the 
Wigner function of any field density matrix, one would 
need to know $\Omega(t)$ as well. To determine this function one takes 
advantage of the experimental setup to measure the Wigner 
function\cite{Lutterbach}, in which a two level atom is prepared in the 
excited state $|e\rangle$, sent trough an array of two low--Q 
cavities R1 and R2 and one high--Q cavity C between them, and is 
detected eventually. The field in C is displaced by the operator 
$D(\alpha) = \exp (\alpha a^\dagger -\alpha^* a)$, by a microwave 
source connected to it. R1 and R2 behave as ``rotation'' operators in 
the Hilbert space of atomic states, $\mid e \rangle \rightarrow (\mid e
\rangle + e^{i\xi} \mid g \rangle) / \sqrt{2}$, and  $\mid g \rangle 
\rightarrow (-e^{-i\xi}\mid e \rangle +  \mid g \rangle) / \sqrt{2}$,
with $\xi=0$ in R1 and $\xi=\pi/2$ in R2. The dispersive atom--field 
interaction in C produces entanglement: The field component associated 
with $\mid e \rangle$ suffers the action of the operator $\exp (i\pi 
(a^\dagger a +1)/2)$, while the one associated with $\mid e \rangle$ 
suffers the action of $\exp (-i\pi a^\dagger a /2)$. It was shown in 
Ref. \cite{Lutterbach} that for this experimental arrangement
\begin{equation}
\Delta P = P_e-P_g = W(-\alpha,-\alpha^*, t) /2,
\end{equation}
where $P_e$ ($P_g$) is the probability to detect the probe atom in the 
upper (lower) state $|e\rangle$ ($|g\rangle$), and $W(-\alpha, 
-\alpha^*, t)$ is the value of the Wigner function corresponding to the
field in thehigh--Q cavity at the time $t$ the probe atom exits this 
cavity. Notice that, for the sake of convenience the normalization of 
$W$ has been changed by a factor of $\pi$.
The Wigner function of the coherent state (\ref{Eq:CState}) is 
\begin{equation}
W(\alpha,\alpha^*,t) = 2 e^{-2(\sigma(t)-\alpha)(\sigma(t)-\alpha)^*}.
\end{equation}
Let's notice that $W(\alpha,\alpha^*,t)$ presents a maximum at $\alpha = 
\sigma(t) = \sigma_0 \exp(-\Lambda(t)-i\Omega(t))$. Since $\Lambda$ can 
be determined from a photocounting experiment, we just need to measure 
$\Omega(t)$. We choose $\alpha=\alpha(t)=\sigma_0 \exp(-\Lambda(t)) 
\exp(-i\Phi(t))$, and adjust $\Phi(t)$ to maximize $\Delta P$. If $\Phi 
(t) = \Omega(t) {\rm mod} (2\pi)$, then we obtain $\Delta P = 1$, 
otherwise, it will be smaller than one, $\Delta P = \exp (-8 \sigma_0^2 
\exp(-2\Lambda) \sin^2((\Omega-\Phi)/2))$. Choosing different exit 
times $t$ we can map $\Omega(t)$. Notice that what we do, in fact, is 
find the right $\alpha(t)$ that brings the field at the cavity C to 
the vacuum. If we succeed in so doing the probe atom is rotated in R1, 
getting in a superposition of upper and lower states, flies free till 
R2 where the rotation is reversed, and goes to the detectors again in 
the upper state.

\section{Concluding Remarks} 
 
In conclusion, we have shown that, if the optical cavities can be 
modelled through the hamiltonian (\ref{Eq:RWAHamiltonian}), then only 
{\it three, experimentally measurable} real functions are necessary to 
characterize their behavior: the mean photon number of an initial
ground 
state, the instantaneous frequency and the rate of change of the number
of excitations. Measuring thes quantities once will enable one to get
the Wigner function for any initial condition. So, provided the
adequacy of the RWA model has been tested the present work provides
for an alternative way to construct the time evolution of Wigner's
functions.

\section{Acknowledgements}

This work was partly funded by FAPESP, CNPq and PRONEX (Brazil), 
and Colciencias, DINAIN (Colombia). K.M.F.R. gratefully acknowledges 
the Instituto de F\'\i sica, Universidade de S\~ao Paulo, for their
hospitality and PRONEX for partial support.

\end{document}